\documentclass[aps, pra, reprint, superscriptaddress, nofootinbib]{revtex4-2}

\usepackage[english]{babel}
\usepackage{import}
\usepackage{enumerate}
\usepackage{graphicx}
\usepackage{color}
\usepackage[dvipsnames]{xcolor}
\usepackage{appendix}
\usepackage[applemac]{inputenc} 
\usepackage{fontenc}  
\usepackage{amsmath}
\usepackage{comment} 
\usepackage{subfigure}
\usepackage{dsfont} 
\usepackage{float}
\usepackage[normalem]{ulem}
\definecolor{linkcolor}{rgb}{0,0,0.6}		
\definecolor{bleu}{HTML}{1732a6}
\usepackage[colorlinks=true, pdfstartview=FitV, linkcolor= linkcolor, citecolor= linkcolor, urlcolor= linkcolor, hyperindex=true, hyperfigures=false]{hyperref}
\newcommand{\ket}[1]{| #1 \rangle}
\newcommand{\bra}[1]{\langle #1 |}
\newcommand{\beq}{\begin{equation}}
\newcommand{\eeq}{\end{equation}}
\newcommand{\bea}{\begin{eqnarray}}
\newcommand{\eea}{\end{eqnarray}}
\newcommand{\ba}{\begin{array}}
\newcommand{\ea}{\end{array}}

\newcommand{\bef}{\begin{figure}}
\newcommand{\eef}{\end{figure}}

\newcommand{\ie}{{\it i.e.\/}\ }
\newcommand{\eg}{{\it e.g.\/}\ }

\newcommand{\setup}[1]{{\textsf{#1}}}

\newcommand{\id}{{{\text{id}}} }
\newcommand{\deph}{{{\text{deph}}} }
\newcommand{\decor}{{{\text{decor}}} }
\newcommand{\reset}{{{\text{therm}}} }

\newcommand{\CI}{{\textsf{C{}}$_\mathrm{x}$}}
\newcommand{\CII}{{\textsf{C}}}
\newcommand{\AI}{{\textsf{A}$_\mathrm{x}$}}
\newcommand{\AII}{{\textsf{A}}}

\usepackage{enumitem}
\setlist{nosep}
\usepackage{bookmark}

\newcommand{\LKB}{{Laboratoire Kastler Brossel, Coll\`ege de France, CNRS, ENS-Universit\'e PSL,
Sorbonne Universit\'e, 11 place Marcelin Berthelot, 75005 Paris, France}}
\newcommand{\JUSSIEU}{{Laboratoire Kastler Brossel, Sorbonne Universit\'e, CNRS, Coll\`ege de France, ENS-Universit\'e PSL, 4 place Jussieu, 75005 Paris, France}}
\newcommand{\LCAR}{{Laboratoire Collisions Agregats R\`eactivit\'e, CNRS, Universit\`e de Toulouse, 118 route de Narbonne, 31062 Toulouse, France}}
\newcommand{\GRENOBLE}{{Universit\'e Grenoble Alpes, CNRS, Grenoble INP, Institut N\'eel, 38000 Grenoble, France}}

\begin{document}

\author{Guillaume C{\oe}uret Cauquil}
\affiliation{\LKB}
\affiliation{\LCAR}

\author{Patrice A. Camati}
\affiliation{\GRENOBLE}

\author{Ir\'en\'ee Frerot}
\affiliation{\GRENOBLE}
\affiliation{\JUSSIEU}

\author{Zheng Tan}
\email{Present address: Innovation Academy for Precision Measurement Science and Technology, The Chinese Academy of Sciences, West No. 30 Xiao Hong Shan, Wuhan 430071, China.}
\affiliation{\LKB}

\author{Alexia Auff\`eves}
\affiliation{\GRENOBLE}
\affiliation{MajuLab, CNRS-UCA-SU-NUS-NTU International Joint Research Laboratory}
\affiliation{Centre for Quantum Technologies, National University of Singapore, 117543 Singapore, Singapore}

\author{Igor Dotsenko}
\email{Contact author: igor.dotsenko@irsamc.ups-tlse.fr}
\affiliation{\LKB}
\affiliation{\LCAR}

\title{Irreversibility of decorrelating processes: an experimental assessment in cavity QED}

\begin{abstract}
Entropy production quantifies the amount of irreversibility of a physical process, leading to fundamental bounds for thermodynamic quantities. It captures the inability to run a physical system forward and then backward, bringing it to the same initial state. Considerable research has been carried out in the last decades to extend entropy production to non-equilibrium quantum processes. We experimentally investigate the entropy production of such forward-backward cycles affected by genuinely quantum irreversibility. Namely, we consider processes realized to erase different types of correlations between two interacting systems, from obliterating solely quantum coherence to completely decorrelating local states. This makes the measurement of entropy production experimentally challenging. Addressing this challenge is the purpose of this paper.
\end{abstract}
\maketitle

\section{Introduction}
The second law of thermodynamics plays a fundamental role in physics establishing the notion of irreversibility and providing constraints on the conversion of heat into work. For thermodynamically open systems, the second law is expressed as the positivity of entropy production~\cite{Zemansky97}, the quantifier of irreversibility. The irreversibility of a physical process can be understood as the inability to bring the system back to its initial state recovering exactly the same resources that were spent in the original process. For example, consider a forward process in which the amount of work $W$ has been extracted from a thermodynamic system that undergoes a transformation from state $A$ to state $B$ while absorbing the heat $Q$ from the environment. This transformation is reversible if and only if there exists a backward process that performs the work $-W$ bringing the system back from $B$ to $A$ while ejecting the heat $-Q$. According to this notion, irreversibility is related to the cost of closing the forward-backward cycle, which corresponds to the amount of energy wasted in the environment. In the classical realm, and for an isothermal cycle, this wasted energy exactly corresponds to the product of the temperature multiplied by the entropy produced.

\begin{figure}[b]
	\begin{center}
    \includegraphics[width=0.8\columnwidth]{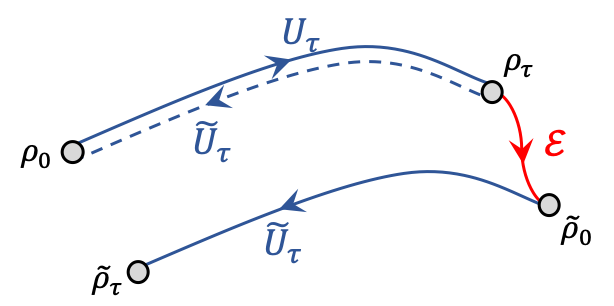}
    \caption{Forward-backward cycle in the presence of an irreversible intermediate process $\mathcal{E}$. In the absence of $\mathcal{E}$, the cycle is completely reversed (dash backward evolution).
    \label{fig:cycle}}
    \end{center}
\end{figure}

In the last decades, considerable progress has been achieved in formally extending the concept of entropy production to microscopic nonequilibrium systems in both the classical~\cite{Esposito_2010,Sagawa_2013,Parrondo2015} and quantum~\cite{Landi2021,Goold_2016,Kosloff_2013} realms. In particular, an influential approach that stimulated a lot of research is the two-point measurement (TPM) scheme in stochastic quantum thermodynamics~\cite{Tasaki2000,Suomela_2014}. In this approach, irreversibility is also associated with closing a forward-backward cycle, but the cycle is modeled with three consecutive processes: The first forward and the last backward processes are unitary, $U_\tau$ and $\tilde U_\tau$, and hence reversible, while the second process, $\mathcal{E}$, is irreversible. The quantum trajectories given by the measured initial and final quantum states, $\rho_0$ and $\tilde\rho_\tau$, respectively, are employed to compute the entropy production of the intermediate irreversible process, see Fig.~\ref{fig:cycle}.

Mathematically, the entropy production takes the form of the Kullback-Leibler-Umegaki (KLU) divergence $D\left(\rho_1||\rho_2\right)$, a common information-theoretic quantifier of the relative distance between a state $\rho_2$ and some reference state $\rho_1$~\cite{Kullback1951,Umegaki1954}. It is defined as $D\left(\rho_1||\rho_2\right)=-\text{Tr}\left[\rho_1\ln\rho_2\right]-S\left(\rho_1\right)$, with $S\left(\rho_1\right)$ being the von Neumann entropy of state $\rho_1$. Crucially, if the matrices $\rho_1$ and $\rho_2$ have reduced rank and different support, then $D$ risks to diverge to infinity. This is especially the case in the quantum realm where irreversibility does not necessarily stem from thermalisation as we shall see shortly. This calls for extra-care in estimating the entropy production from the measurement data. Exploring these experimental challenges of quantum thermodynamics is the main purpose of this paper.

The divergence $D$, as the entropy $S$, is not associated with any observable. It is a nonlinear function of the states, rendering intricate experimental assessment of the entropy production. Nonetheless, the entropy production has been experimentally deduced not only for the typical thermodynamic process of thermalization~\cite{Batalhao2015,Camati2016,Najera-Santos2020}, but also for genuinely quantum processes such as quantum measurements~\cite{Mancino2018,Harrington2019,Rossi2020}. In this paper, we experimentally assess the entropy production of several different irreversible processes that partially or completely erase the correlations of a bipartite system~\cite{Landi2021}.

Entanglement, or, more generally, quantum correlations~\cite{Adesso2016}, has been shown to be a major resource for a plethora of information processing tasks~\cite{Galindo2002,Gisin2002}. However, it is very fragile and easily degraded by ubiquitous processes such as decoherence and thermalization of the constituting quantum systems. We focus on three distinct decorrelating processes that can act on a bipartite quantum system: decoherence, complete decorrelation, and local thermalization. The decoherence process completely eliminates the coherence of either one or both subsystems in their eigenbasis. This process erases all quantum correlations, but retains classical correlations between the two systems. Complete decorrelation erases all correlations, classical and quantum, that might be present in the bipartite state, leaving the reduced states intact. Local thermalization brings one of the subsystems to a thermal state, which also completely eliminates all correlations in the process. For these processes, entropy production has been shown to quantify the entropic cost of erasing the corresponding correlations~\cite{Landi2021}. For the local thermalization process, the entropy production further accounts for the irreversibility of changing the reduced state to a thermal state.

We experimentally study entropy production in a test bipartite system composed of a two-level atom (circular Rydberg) resonantly coupled to the electromagnetic field of a microwave cavity (high-quality superconducting)~\cite{Haroche13}. In order to assess the entropy production, we have to experimentally estimate quantum states at reference points of the realized cycle. A typical reconstruction method is based on the maximum likelihood estimation (MLE) of quantum states from measured data. The state maximizing the data likelihood may not have full support on the Hilbert space, a feature of the MLE that is usually innocuous and therefore ignored. However, a reference state of reduced support used for computing the divergence may lead to potentially spurious nonphysical divergences, preventing us from adequately computing the physical entropy. To avoid this issue, we adapt the MLE protocol proposed in Ref.~\cite{Six2016} and obtain more realistic full rank density operators estimating quantum states in the cycle. 

We present our results in the following way. We revisit the theory for obtaining the entropy production in the considered decorrelating irreversible processes via the TPM scheme in Sec.~\ref{sec:theory}. Section \ref{sec:experiment} presents the experimental setup adapted to simulate the action of different types of environments upon a quantum bipartite system. The methodology and experimental details to assess and estimate the entropy production are introduced in Sec.~\ref{sec:reconstruction}. Experimental simulation of irreversible processes along with the reconstructed system's states at different phases are presented in Sec.~\ref{sec:environments}. Finally, we conclude our work in Sec.~\ref{sec:conclusions}.

\section{Two-point measurement}\label{sec:theory}

Entropy production quantifies the amount of irreversibility of a physical process. Employing TPM schemes~\cite{Manzano2018,Landi2021}, a number of fluctuation relations~\cite{Morikuni2011,Campisi2011,Murashita2017} for thermodynamic stochastic variables have been derived for a variety of physical processes. The underlying physical protocol considered in such methods can be decomposed into three distinct processes: a unitary forward process $U$, an intermediate (irreversible) process $\mathcal{E}$, and a unitary backward process $\Tilde{U}$, as schematically shown in Fig.~\ref{fig:cycle}. In the following, we recapitulate the main ingredients of the TPM scheme and apply it to the decorrelating processes we considered.

After the forward, and reversible, process is realized, the system evolves according to an uncontrollable dynamics represented by the CPTP map $\mathcal{E}$, bringing the state $\rho_{\tau}$ to a new state $\tilde{\rho}_{0}=\mathcal{E}\left(\rho_{\tau}\right)$. The lack of control over this intermediate process means that the suitable reverse evolution of the controllable forward process will not be capable of bringing the system back to its original state $\rho_{0}$. Such an inability stems from the irreversibility  of the intermediate process. Therefore, it is during this intermediate irreversible process that the entropy production arises. This process has often been taken as complete thermalization, leading to well-established results. This captures in particular the famous Jarzynski's protocol, where a quantum system is driven out of thermal equilibrium and let relax before being brought back to its initial state. However, other types of irreversible processes can be considered, like decorrelating processes erasing completely or partially the correlations in a bipartite system~\cite{Landi2021}. We come back to this class of processes shortly. 

After the irreversible intermediate stage, the backward process attempts to bring the system back to its initial state, \ie the initial state $\rho_{0}$ of the forward process. The backward evolution is performed via the controllable Hamiltonian $H_{\text{B}}\left(t\right)=H_{\text{F}}\left(\tau-t\right)$ bringing the initial backward state $\tilde{\rho}_{0}$ to $\tilde{\rho}_{\tau}=\tilde{U}_{\tau,0}\tilde{\rho}_{0}\tilde{U}_{\tau,0}^{\dagger}$, where $\tilde{U}_{\tau,0}=\mathcal{T}\exp\left\{ -\frac{i}{\hbar}\int_{0}^{\tau}dt\:H_{\text{B}}\left(t\right)\right\} $. We note that $\tilde{U}_{\tau,0}=U_{\tau,0}^{\dagger}$~\cite{Camati2018}. Being fully controllable, the backward process is also reversible, hence does not carry any entropy production. 

In the TPM scheme, a measurement is performed at the beginning and at the end of the forward and backward processes. Each pair of measurement outcomes defines a forward and a backward trajectory, respectively. The entropy production $\left\langle \Sigma\right\rangle$ of the intermediate process is obtained by appropriately averaging the ratio of forward and backward trajectory probabilities. The resulting entropy production ends up being given by the KLU divergence:
\begin{equation}
\left\langle \Sigma\right\rangle =D\left(\rho_{\tau}||\tilde{\rho}_{0}\right)=D\left(\rho_{0}||\tilde{\rho}_{\tau}\right).
\end{equation}
The last equality states that the entropy production in the irreversible process is given by the divergence between the initial and final states of this process. Similarly, since the forward and backward processes are reversible, \ie have no entropy production, $\left\langle \Sigma\right\rangle$ also equals the divergence between the initial and final states of the whole protocol. 
 
If the intermediate process was given by the identity operation $\tilde{\rho}_{0} = \mathcal{E}_{\id}\left(\rho_{\tau}\right)=\rho_{\tau}$, then the entropy production would be $\left\langle \Sigma_{\id}\right\rangle =0$, in agreement with the notion that this process is reversible. On the other hand, if we consider the complete thermalization process, the CPTP map is given by $\tilde{\rho}_{0} = \mathcal{E}_{\text{therm}}\left(\rho_{\tau}\right)=\zeta_{\beta}$, where $\zeta_{\beta}=\exp(-\beta\left[H_{\text{F}}\left(\tau\right)-F_{\beta}\left(\tau\right)\right])$ is the Gibbs state at inverse temperature $\beta$ and with Hamiltonian $H_{\text{F}}\left(\tau\right)$, and the equilibrium Helmholtz free energy is $F_{\beta}\left(\tau\right)=-\beta^{-1}\ln(\text{Tr}\left[\zeta_{\beta}\right])$. The entropy production of such intermediate thermalization process becomes $\left\langle \Sigma_{\text{therm}}\right\rangle =\Delta S_{\text{therm}}-\beta Q_{\text{therm}}$~\cite{Landi2021}, where $\Delta S_{\text{therm}}$ and $Q_{\text{therm}}$ are the entropy and energy change of the system during thermalization. This can be seen as the quantum expression of the famous Clausius inequality. The identity and thermalization processes are somewhat typical thermodynamic processes: either one does nothing, in which case there is no irreversibility, or one lets the system thermalize with a heat reservoir, an ubiquitous process in thermodynamics. Now, we discuss the entropy production of decorrelating processes, capable of erasing classical or quantum correlations even without the presence of heat reservoirs.

In the following, we assume that the system is composed of two subsystems, $A$ and $B$, and that the unitary evolution results from their interaction. Let us start considering a local-dephasing process acting on both subsystems, defined by the CPTP map
\begin{eqnarray}
    \mathcal{E}_{\deph}\left(\rho^{AB}\right)&=&\sum_{k,l}P_{k}^{A}P_{l}^{B}\rho^{AB}P_{k}^{A}P_{l}^{B}\nonumber\\
    &=&\sum_{k,l}p\left(k,l\right)P_{k}^{A}P_{l}^{B},
    \label{eq:CPTP_map_dephasing}
\end{eqnarray}
where $\left\{ P_{k}^{A}\right\}$ and $\left\{ P_{k}^{B}\right\}$ are the eigenoperators of the reduced states $\rho^{A}$ and $\rho^{B}$, respectively, and $p\left(k,l\right)=\text{Tr}_{AB}\left[P_{k}^{A}P_{l}^{B}\rho^{AB}\right]$. The resulting state of Eq.~(\ref{eq:CPTP_map_dephasing}) is a classically-correlated quantum state~\cite{Bellomo2014}, which means that the local-dephasing process erases all the quantum correlations present in the state $\rho^{AB}$. Physically,  local dephasing can be performed in two ways. Firstly, each subsystem $A$ and $B$ is independently subject to a decoherence process in their eigenbasis. Secondly, a non-selective projective measurement in the eigenbasis of $\rho^{A}$ and $\rho^{B}$ is jointly realized, the state in Eq. (\ref{eq:CPTP_map_dephasing}) being the resulting non-selective state. In the latter situation, $p\left(k,l\right)$ is a joint probability of obtaining outcomes $k$ and $l$ for the joint measurement. 

Generally, starting from a separable state of the systems $A$ and $B$, the forward process will create correlations between them, which are quantified by the mutual information 
$I\left(\rho^{AB}\right)=D\left(\rho^{AB}||\rho^{A}\otimes\rho^{B}\right)$. When the local-dephasing evolution $\mathcal{E}_{\deph}$ is considered as the irreversible intermediate process within the TPM scheme, its resulting entropy production is
\begin{align}
    \left\langle \Sigma_{\deph}\right\rangle = & D\left(\rho_{\tau}^{AB}||\mathcal{E}_{\deph}\left(\rho_{\tau}^{AB}\right)\right)\nonumber \\
    = & S\left(\mathcal{E}_{\deph}\left(\rho_{\tau}^{AB}\right)\right)-S\left(\rho_{\tau}^{AB}\right).
\end{align}
The last expression corresponds to the so-called relative entropy of coherence~\cite{Baumgratz2014}, which is a measure of the quantum correlations present in the state $\rho_{\tau}^{AB}$~\cite{Luo2008}. Alternatively, we can write 
\begin{equation}
    \left\langle \Sigma_{\deph}\right\rangle = I\left(\rho_{\tau}^{AB}\right)-I\left(\mathcal{E}_{\deph}\left(\rho_{\tau}^{AB}\right)\right), 
\end{equation} 
illustrating the fact that local dephasing erases only the quantum correlations: $I\left(\rho_{\tau}^{AB}\right)$ quantifies the total amount of correlations of $\rho_{\tau}^{AB}$, both classical and quantum, while $I\left(\mathcal{E}_{\deph}\left(\rho_{\tau}^{AB}\right)\right)$ quantifies the correlations of $\mathcal{E}_{\deph}\left(\rho_{\tau}^{AB}\right)$, containing only classical correlations.

In order to connect with the experimental implementation discussed in Sec.~\ref{sec:environments}, consider the similar dephasing process in which only system $B$ undergoes the decoherence process,
\begin{equation}
\mathcal{E}_{\text{deph,B}}\left(\rho^{AB}\right)=\sum_{l}P_{l}^{B}\rho^{AB}P_{l}^{B}=\sum_{l}p\left(l\right)\rho_{l}^{A}\otimes P_{l}^{B},\label{eq:CPTP_map_single_dephasing}
\end{equation}
where
\begin{equation}
\rho_{l}^{A}=\frac{\text{Tr}_{B}\left[P_{l}^{B}\rho^{AB}\right]}{p\left(l\right)}=\sum_{k}p\left(k|l\right)P_{k|l}^{A}
\end{equation}
and $P_{k|l}^{A}$ are the eigenprojectors of state $\rho_{l}^{A}$. 
 If all states $\rho_{l}^{A}$ are diagonalized by the same set of projectors $\left\{ P_{k}^{A}\right\} $, the resulting state of Eq.~(\ref{eq:CPTP_map_single_dephasing}) coincides with the resulting state of Eq.~(\ref{eq:CPTP_map_dephasing}), with $p\left(k,l\right)=p\left(k|l\right)p\left(l\right)$. In this particular case, we have $\mathcal{E}_{\text{deph,B}}\left(\rho^{AB}\right)=\mathcal{E}_{\deph}\left(\rho^{AB}\right)$ meaning that all the quantum correlations of the state $\rho^{AB}$ have been erased by a single dephasing in system $B$. Therefore, the local dephasing process on system $B$ has the same effect of the local dephasing process for both $A$ and $B$. In our experimental implementation, we consider only the dephasing in one subsystem.

Let us now consider the complete decorrelating process, where both quantum and classical correlations are fully erased. For an arbitrary state $\rho^{AB}$, the map describing this process reads $\mathcal{E}_{\decor}\left(\rho^{AB}\right)=\rho^{A}\otimes\rho^{B}$. This map is generally non-linear, since $\rho^{AB}$ appears inside $\rho^{A}$ and $\rho^{B}$, and therefore it does not correspond to a CPTP map~\cite{Terno1999} and cannot be realized physically. However, such a map can be simulated using simultaneously two copies of the same system and letting interact subsystems of different copies, as will be presented in Sec.~\ref{sec:environments}. Assuming its implementation, the resulting entropy production from the TPM scheme gives
\begin{equation}
\left\langle \Sigma_{\decor}\right\rangle =I\left(\rho_{\tau}^{AB}\right),
\end{equation}
demonstrating that the entropy production of the complete decorrelating process is precisely given by the information-theoretic cost of erasing all the correlations.

The local thermalization of a bipartite system also comprises a decorrelating process that furthermore involves the typical thermal environment. After the forward process, the system $B$ is coupled to a thermal reservoir that locally thermalizes the system $B$ alone. The CPTP map associated with this process maps the arbitrary bipartite state as $\mathcal{E}_{\text{loc-therm}}\left(\rho^{AB}\right)=\rho^{A}\otimes\zeta_{\beta}^{B}$, where $\rho^{A}$ is simply the reduced state of system $A$ while $\zeta_{\beta}^{B}$ is the Gibbs state reached by system $B$. During the thermalization of system $B$, the correlations present in $\rho^{AB}$ are completely erased. Therefore, the resulting entropy production is given by the addition of the complete de-correlation process and the thermalization process,
\begin{equation}
\left\langle \Sigma_{\text{loc-therm}}\right\rangle = I(\rho_\tau^{AB})+D\left(\rho^{B}||\zeta_{\beta}^{B}\right).
\end{equation}
In the context of the quantum information processing, deterministic preparation of the system in a given initial state, that is not necessarily thermal from the classical point of view, can also be seen as thermalization with a 'state-preparation' reservoir. In the experimental protocols realized in the present work, the role of local thermalization is played by the reset process bringing one of the subsystems into its initial pure state, see Sec.~\ref{sec:environments}.

\section{Experimental system}\label{sec:experiment}

One of the simplest systems for studying irreversibility is composed of a single two-level atom (qubit) interacting with a resonant cavity (harmonic oscillator). This model features quantum properties in a large Hilbert space while staying relatively simple and providing a good understanding and description of environment interventions. We choose the resonant atom-cavity interaction as a reversible evolution $U$ allowing energy and information exchange within the system, relevant for thermodynamical analysis. Finally, we set the interaction duration to a quarter of the Rabi oscillation period (so-called $\pi/2$ pulse). For the initially excited qubit and the resonator in its ground state, this interaction prepares maximally entangled qubit-resonator states, for which decorrelation effects are the most visible and, thus, easier to detect. In fact, a mutual information between maximally entangled qubit and oscillator reaches its maximal value of 2 bits, insuring a maximal contrast between correlated and decorrelated states.

\begin{figure}[t]
	\begin{center}
    \includegraphics[width=0.99\columnwidth]{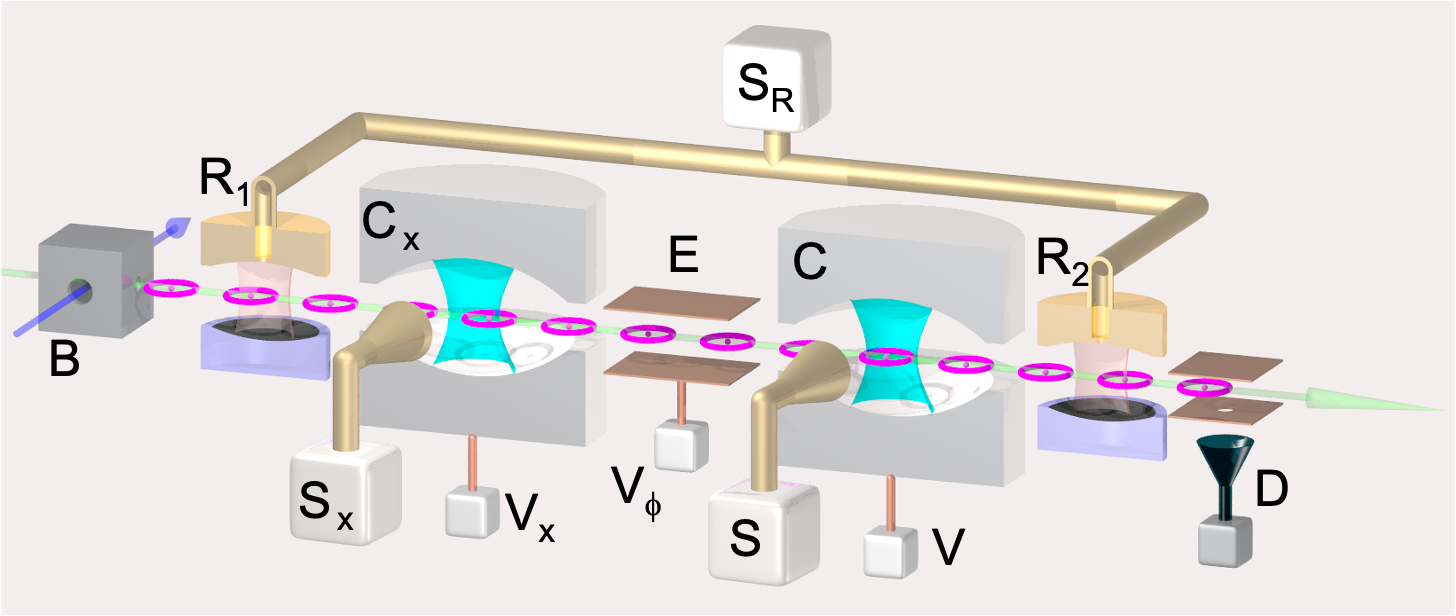}
    \caption{Scheme of the experimental setup. Flying circular Rydberg atoms (toroids), excited in box \setup{B}, can sequentially interact with two high-quality microwave cavities, \CI{} and \CII{}. The atom-cavity detuning is controlled by electric field applied by voltage potential \setup{V}$_\mathrm{x}$ and \setup{V}, respectively. The atomic state is manipulated by a low-quality cavity \setup{R}$_1$ and is measured by ionization detector \setup{D}. The cavity state is reconstructed using Ramsey interferometer (zones \setup{R}$_1$ and \setup{R}$_2$ fed by source \setup{S}$_\setup{R}$) with a sequence of probe atoms and homodyne microwave field injected by sources \setup{S} and \setup{S}$_\mathrm{x}$.
    \label{fig:setup}}
    \end{center}
\end{figure}

We realize the model qubit-resonator system in a cavity QED setup \cite{Haroche13,Najera-Santos2020}, schematically presented in Fig.~\ref{fig:setup}. Qubits are encoded into circular Rydberg atoms with principal quantum numbers $n=50$ and $51$. The corresponding levels, denoted by $\ket{g}$ and $\ket{e}$, are separated by 51~GHz. The resonator is a high-quality microwave superconducting cavity $\setup{C}$. Its photon lifetime of $65$ ms is much larger than the overall duration of the experiment. 

Rubidium-85 atoms are effusing from an oven, velocity selected by Doppler effect (250 m/s) and excited in box \setup{B} to the circular Rydberg state $\ket{e}$. The excitation is realized by a series of laser, microwave and radio-frequency pulses. The atomic states can be additionally manipulated by classical fields in two Ramsey zones, $\setup{R}_1$ and  $\setup{R}_2$, fed by a source $\setup{S}_\setup{R}$. The atom-cavity detuning is controlled by an electric field, applied across the metallic resonator's mirrors with potential $\setup{V}$, by means of the quadratic Stark effect. In this way, we set the exact interaction time and realize the resonant $\pi/2$ interaction. Finally, atomic states are detected by ionization in a channeltron detector \setup{D}. The random atom number per atomic sample obeys Poisson statistics. In order to ensure a single-atom interaction, we reduce the mean atom number per sample to about $0.2$ atoms and post-select experimental results on detection of exactly one atom.

Our experimental setup consists of two similar cavities, \CI{} and \CII{}, with independent control over atom-cavity interaction. All experiments are performed with cavity \CII{}. The auxiliary cavity \CI{} is used in addition to \CII{} in the realization of the complete decorrelation environment and local thermalization, as described later. 

We perform full quantum state tomography to access quantum state, \ie a complete density matrix, of the joint atom-cavity system. The cavity field is probed by quantum non-demolition (QND) measurement by means of a long sequence of atoms, that interact dispersively with the cavity field and detected one by one. Being measured in the Ramsey interferometer configuration, composed of two Ramsey cavities, $\setup{R}_1$ and $\setup{R}_2$ installed before and after \CII{}, they provide us information on the photon number distribution in \CII{} \cite{Guerlin07}. The information on field coherence is extracted by additional homodyne injection  of a reference microwave field from source $\setup{S}_{}$ into \CII{} \cite{Deleglise08}. The atomic states in energy basis are directly detected by ionization in \setup{D}. An additional $\pi/2$ Rabi pulse applied in zone $\setup{R}_2$ before ionization allows for qubit measurement in the $x$ basis of the Bloch sphere. All different measurement settings and the corresponding measurement results from cavity probing and atom detection are combined together in a single MLE algorithm \cite{Metillon19} providing us a reconstructed atom-cavity state of the composite system.

\section{Quantum state reconstruction}\label{sec:reconstruction}
Estimation of the system's density matrix from an experimental dataset is based on the algorithm presented in Ref.~\cite{Metillon19}. With sufficiently sensitive measurements and sufficient data, it provides without any particular difficulty the state estimate $\rho_{\textrm{ML}}$ with maximal likelihood of the measured data. Any observable can be directly computed from the set of all measurement results without the need to calculate it from $\rho_{\textrm{ML}}$. As for other functions of the density matrix, they can be computed from $\rho_{\textrm{ML}}$. The problem arises for non-linear functions with divergence tendency, like relative entropy $D\left(\rho_1||\rho_2\right)$. If computed between two matrices with different non-full support, the resulting function can be very sensitive to any minor noise in the state reconstruction procedure, making its further use pointless. In the following, we discuss the difficulties of managing the relative entropy for states without full support as well as our method to circumvent this issue. 

\subsection*{Estimating the density matrix from experimental dataset: a Bayesian framework}
Our experimental method for reconstructing quantum states is described in details in Ref.~\cite{Six2016}. Here we summarize the key points relevant to the estimation of thermodynamical quantities. To acquire information on a quantum state to reconstruct, the preparation-measurement sequence is repeated $N$ times for each of the decorrelating processes described above (typically, $N\sim 10^4-10^5$). The possible measurement outcomes (sequences of detected atomic states after the atom-cavity interaction), labeled as $i$, are mathematically described by so-called effect matrices $E_i$, used in Ref.~\cite{Gammelmark2013} to extract maximum information on a measured quantum trajectory. They are positive semidefinite matrices and $\sum_i E_i = \mathds{1}$. If the system's density matrix is $\rho$, then outcome $i$ is obtained with probability $p(i|\rho) = {\rm Tr}(\rho E_i)$. The key issue is to estimate $\rho$ from the sequence of measurement outcomes ${\cal D} = (i_1, \dots, i_N)$. Notice that, as long as the experimental runs are independent from each other, we have $p({\cal D}|\rho) = \prod_i p(i|\rho)$. We adopt here a Bayesian point of view, in which, given the experimental dataset ${\cal D}$, we attribute a likelihood $L$ to every density matrix using
\begin{equation}
    L(\rho |{\cal D}) \propto  p({\cal D}|\rho).
\end{equation}
The likelihood function is the basic mathematical object from which physical properties of the system are then estimated, as well as their corresponding uncertainties.

\subsection*{Estimating functions of the density matrix: mean Bayesian estimate}
In this work, we focus on estimating non-linear functions of the density matrices, such as the entropy and the KLU divergence between two density matrices. Formally, for a given function $f(\rho)$, we introduce its corresponding likelihood:
\begin{equation}
    L_f(x|{\cal D}) \equiv \int d\rho ~L(\rho|{\cal D})~ \delta(x - f(\rho)),
\end{equation}
where the integral is over all density matrices (using Haar measure) and $\delta$ is the Dirac function. From the $L_f$ function, we may then propose an estimate for $f(\rho)$ given the experimental dataset, as well as its corresponding uncertainty. Here, as an estimate we simply take the average of the $L_f$ distribution:
\begin{eqnarray}
    f_{\rm est} &=& \langle x \rangle_{L_f} \\
    &=& \int dx ~x~L_f(x|{\cal D}) \\
   &=& \int d\rho ~f(\rho) ~L(\rho|{\cal D}) ~. \label{eq_fest}
\end{eqnarray}
Intuitively, every density matrix which is not excluded by the experimental results contributes to the estimate of $f$, weighted by its corresponding likelihood $L(\rho |{\cal D})$.

Similarly, the uncertainty on this estimate is evaluated as the standard deviation of the $L_f$ distribution:
\begin{equation}
    \delta f_{\rm est} = \sqrt{\langle x^2 \rangle_{L_f} - \langle x \rangle_{L_f}^2} ~.\label{eq_errfest}
\end{equation}
We emphasize that the above-described procedure is not equivalent to the widely-used maximum-likelihood estimate, in which one first introduces the density matrix $\rho_{\rm ML} = {\rm argmax}_\rho [L(\rho|{\cal D})]$ and then uses $f_{\rm est} = f(\rho_{\rm ML})$. While both approaches converge to the same estimate in the limit of infinitely many samples ($N\to \infty$), they lead to very different results when estimating non-linear functions with a finite number of samples. In particular, it is common that $\rho_{\rm ML}$ has some zero eigenvalues, \ie is rank deficient, which leads to nonphysical divergences when estimating the KLU divergence between two density matrices. Furthermore, the approach implemented in this work is conceptually more satisfactory than the MLE reconstruction, as it incorporates the complete statistical information contained in the likelihood function, and is not plagued by the possibly very atypical properties of $\rho_{\rm ML}$. Notice also that our approach is conceptually very similar to the so-called mean Bayesian estimate proposed in Ref.~\cite{Blume-Kohout_2010}, where one introduces the density matrix $\rho_{\rm MBE} = \int d\rho ~\rho~L(\rho|{\cal D})$, and then uses $f_{\rm est} = f(\rho_{\rm MBE})$. While this is equivalent to our approach when $f$ is a linear function of $\rho$, our estimate is more generally suited for estimating arbitrary functions of the density matrix.

\subsection*{Sampling density matrices according to their likelihood: Monte-Carlo approach}
We limit the cavity's Hilbert space size to $4$, allowing thus states with up to 3 photons, reasonable for interactions and protocols realized in this work. Given that density matrices of the atom-cavity system are $8 \times 8$ hermitian matrices, one cannot perform directly the integration in a high-dimensional state space as required by Eq.~\eqref{eq_fest} and \eqref{eq_errfest}. We circumvent this problem by implementing a numerical Monte-Carlo sampling over density matrices. Concretely, we sample an ensemble of $n$ density matrices $(\rho_1, \dots, \rho_n)$ according to their likelihood, such that $(1/n)\sum_{k=1}^n f(\rho_k) =  \int d\rho ~f(\rho) ~L(\rho|{\cal D}) + O(1/\sqrt{n})$. Namely, averaging over the $n$ sampled density matrices is equivalent to averaging over the $L$ distribution, up to $1/\sqrt{n}$ corrections. This is achieved using the Metropolis algorithm as detailed in Appendix \ref{app_MonteCarlo}.

\section{Simulation of environment}\label{sec:environments}

In the following, we present the experimental realization of several environments considered in Ref.~\cite{Landi2021} and discussed in Sec.~\ref{sec:theory}. In order to make the environment effect more pronounced, we choose a resonant $\pi/2$ interaction between the atom and the cavity as a unitary forward evolution $U_{\text{F}}$. The experimental cycle starts with an excited atom in state $\ket{e}$ and the empty cavity in the vacuum state $\ket{0}$, with the corresponding initial joint state 
\begin{equation}\label{eq:rho_0}
    \rho_0 = \ket{e0}\bra{e0}.
\end{equation}
The forward evolution $U_{\text{F}}$ transforms $\rho_0$ into a maximally entangled atom-cavity state 
\begin{equation}\label{eq:rho_tau}
    \rho_\tau = \frac{1}{2}\left(\ket{e0}+\ket{g1}\right) \left(\bra{e0}+\bra{g1}\right).
\end{equation}
This state exhibits the maximum mutual information $I$ of 2 bits, which is the limit for a two-level subsystem. 

To close the cycle, we perform the backward unitary evolution $U_{\text{B}}=U_{\text{F}}^\dagger$. It starts by first applying the $\sigma_x$ operation on the atomic qubit state and then activating the resonant $\pi/2$ interaction as in $U_{\text{F}}$. Note that this protocol is similar to the spin-echo technique. Since this cycle must be completed while the atom is crossing the spatial Gaussian mode of the cavity, we realize the first resonant interaction $U_{\text{F}}$ before the atom reaches the center of the cavity. The $\sigma_x$ operation is applied when the atom is at the center by detuning it from the cavity resonance by 1~MHz during 0.5~$\mu$s. The exact $\pi$ spin-flip is calibrated using Ramsey interferometer \cite{Penasa16}. Finally, the second resonant interaction included in $U_{\text{B}}$ is applied while the atom is crossing the second half of the cavity mode. 

\begin{figure}[t]
	\begin{center}
    \includegraphics[width=1\columnwidth]{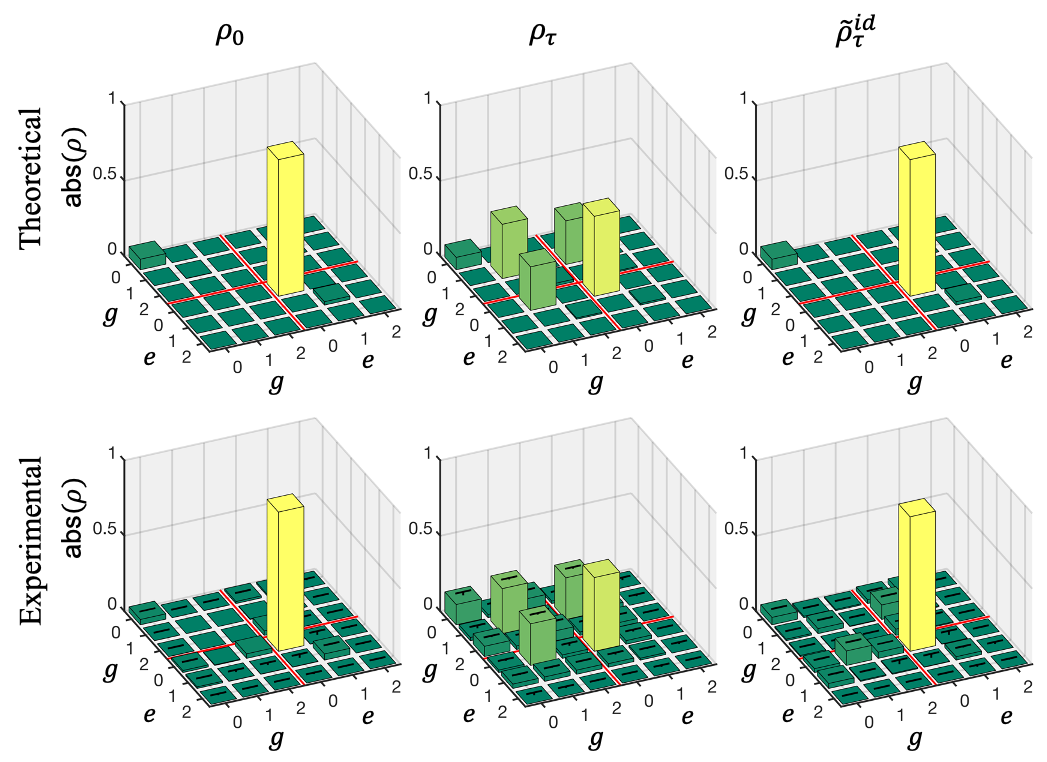}
    \caption{Reconstructed atom-cavity states. Three columns correspond to initial $\rho_0$, intermediate entangled $\rho_\tau$ and final $\tilde\rho_\tau$ states. Upper row: theoretical states based on known experimental imperfections. Lower row: MLE reconstruction. For each atomic state, the cavity photon number is limited by 3 photons. For the sake of clarity, only the absolute values of the density operators are plotted and only up to 2 photons. Red lines on the matrix base delimit the atomic states $e$ and~$g$. 
    \label{fig:states}}
    \end{center}
\end{figure}

Figure \ref{fig:states} shows the reconstructed atom-cavity states (absolute values) with the MLE error bars. Here and in the following, theoretical states account for known experimental imperfections. The main effects include finite atomic and photon lifetimes (nonzero population in $\ket{g0}$), dispersion in atom-cavity coupling (reduced coherences $\ket{e0}\bra{g1}$), non-ideal $\sigma_x$ flip (residual coherences in $\tilde \rho^{\id}_\tau$), and non-negligible presence of a second non-detected atom (populations with double excitation $\ket{e1}$). 

The off-diagonal elements of the entangled state $\rho_\tau$ represent quantum coherence, while the classical correlation between the sub-systems is encoded into the diagonal elements of its joint state $\rho_\tau$. Mutual information of the experimental $\rho_\tau$ is $I(\rho_\tau)=0.88$ bits, that is lower than its ideal value of $2$ bits. It sets the upper limit on possible information erasure by any environment.

\subsection*{No environment action}

The first situation we want to observe involves no decorrelation process, \ie no environment coupling. It is equivalent to the forward-backward cycle composed only of $U_{\text{F}}$ and $U_{\text{B}}$. Ideally, this cycle brings the system back to its initial state, realizing the identity operation:  
\begin{equation}\label{eq:rho_tau_tilde}
  \tilde\rho_{\tau}^{\id}=U_{\text{F}}^{\dagger}U_{\text{F}}\rho_{0}U_{\text{F}}^{\dagger}U_{\text{F}} = \rho_{0}.
\end{equation}
The ideal entropy production for this identity process is zero, 
$\left\langle \Sigma_{\id}\right\rangle =D\left(\rho_{0}||\tilde\rho_{\tau}^{\id}\right) = 0$. 

\begin{figure}[t]
	\begin{center}
    \includegraphics[width=0.99\columnwidth]{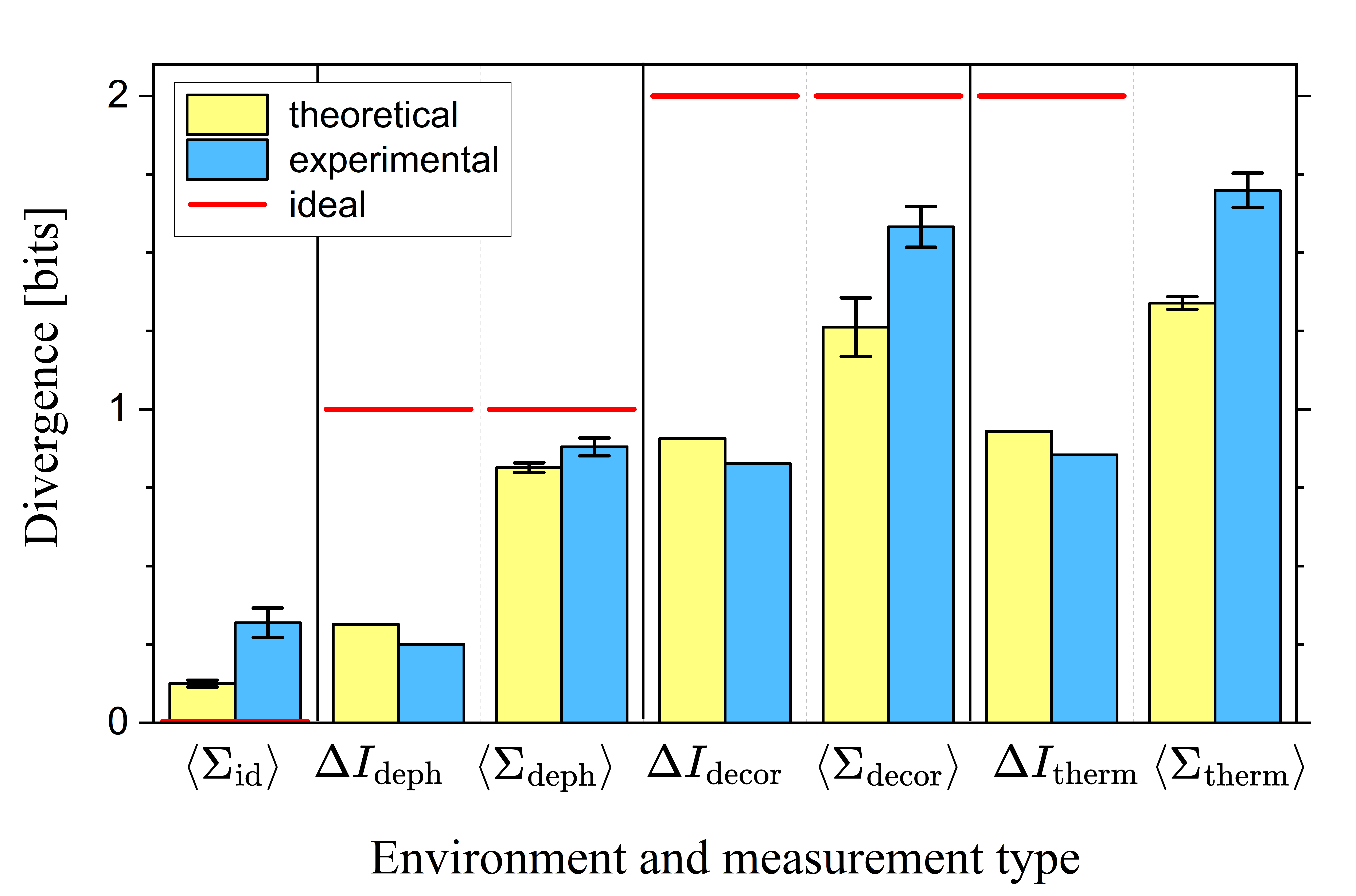}
    \caption{Entropy production, $\langle\Sigma\rangle$, and change of mutual information, $\Delta I$, for different environments. Theoretical results include numerical simulation of both the experimental setup with known imperfections and the quantum state reconstruction procedure. Uncertainty of $\langle\Sigma\rangle$ are obtained from the Monte-Carlo sampling of density matrices, see Appendix~\ref{app_MonteCarlo}. Information erasure $\Delta I$ is computed directly from the MLE density operators. Red lines indicate the ideal values in the absence of any experimental imperfection and reconstruction noise. The ideal value of $\left\langle \Sigma_{\reset}\right\rangle$ is infinite and, thus, not shown on the graph.
    \label{fig:results}}
    \end{center}
\end{figure}

Figure~\ref{fig:results} shows ideal, numerically simulated and experimentally measured entropy production for different  types of environments. Ideal values of information erasure, based on discussion in Sec.~\ref{sec:theory} applied to our atom-cavity system interacting resonantly, are indicated by red lines. For the no-environment case (\ie identity process), both simulated and measured $\left\langle \Sigma_{\id}\right\rangle$ are close to each other, but stay larger than the ideal zero value. Since the divergence $D$ is limited to non-negative values only, any noise in the system's state due to technical imperfections in the state preparation or reconstruction increases the value of the real $\langle \Sigma_{\id}\rangle$. Its value thus quantifies the quality of the realized experimental sequence and its unitary operations, as well as of the MLE protocol. 

It is important to note that, here and in the following, all $\langle\Sigma\rangle$ and their uncertainty are computed with Monte-Carlo sampling of density matrices, as explained in Sec.~\ref{sec:reconstruction} and Appendix~\ref{app_MonteCarlo}. If computing these quantities directly from the reconstructed states presented in Figs.~\ref{fig:states} and \ref{fig:states2}, the difference between their ideal, numerically simulated and experimentally measured values is much larger than that seen in Fig.~\ref{fig:results} and is very sensitive to experimental uncertainties and MLE parameters. Thus, a significant part of the presented work was dedicated to the development of the more careful data treatment providing more realistic states and their non-linear functions.

\subsection*{Local dephasing}
All quantum correlations between two subsystems can be erased by applying local dephasing to one of them. We realize this process by setting large detuning between atomic and cavity frequencies of several~MHz for $0.8$~$\mu$s, which is done by applying a large electric field \setup{V} to the atom across \CII{}. Since the exact electric field value fluctuates, it induces a random phase shift accumulated between two subsystems, hence dephasing the system. The efficiency of the phase erasure is verified by applying this detuning in a Ramsey interferometry sequence, where it completely suppresses any contrast on the Ramsey fringes. The ideal dephased state takes the form of a statistical mixture
\begin{equation}
  \tilde\rho_{0}^{\deph} = \frac{1}{2}\left(\ket{e0}\bra{e0}+\ket{g1}\bra{g1}\right).
\end{equation}
The further process $U_\textrm{B}$ does not change this decohered mixture, making $\tilde\rho_{\tau}^{\deph}=\tilde\rho_{0}^{\deph}$. 

\begin{figure*}[t!]
    \includegraphics[width=1.00\textwidth]{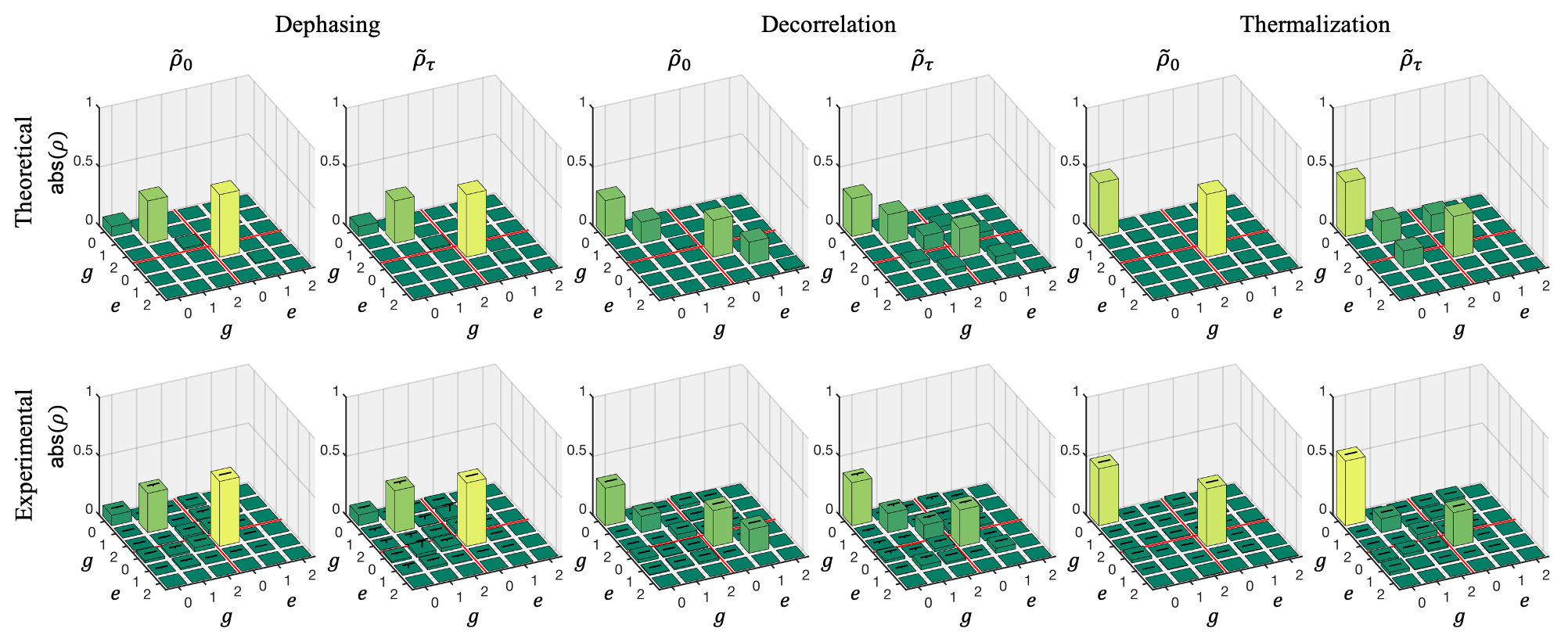}
    \caption{Reconstructed atom-cavity states for three types of decorrelating processes: dephasing, complete decorrelation and thermalization of the cavity to its initial state. Upper and lower rows are theoretical (numerically simulated) and experimentally reconstructed density matrices.
    \label{fig:states2}}
\end{figure*}

Figure \ref{fig:states2} presents the theoretical and experimental states $\tilde\rho_0$ and $\tilde\rho_\tau$ for this process. As expected, the state $\tilde\rho_0$ after the dephasing (Fig.~\ref{fig:states2}) is close to the diagonal elements of the initial entangled state $\rho_\tau$ (Fig.~\ref{fig:states}) that are left after erasing all quantum coherence. The experimentally implemented dephasing protocol is thus close to optimal.
The corresponding $\left\langle \Sigma_{\deph}\right\rangle$ together with the change of the mutual information $\Delta I_{\deph} = I(\rho_\tau)-I(\tilde\rho_0)$ 
is plotted in Fig.~\ref{fig:results}. 

Although we expect the values for $\left\langle \Sigma_{\deph}\right\rangle$ and $\Delta I_{\deph}$ to be identical, the observed entropy production for the full cycle, it is found to be higher than the amount of erased correlations in both simulation and experiment. This is likely due to the imperfection of the forward and backward processes, which are not perfectly unitary and inverse of each other. In this case, additional entropy can be produced. Finally, since the experimentally measured values of $\left\langle \Sigma_{\deph}\right\rangle$ and $\Delta I_{\deph}$ are close to the simulated ones, the experimental atom-cavity system is well understood.

\subsection*{Complete decorrelation}

\begin{figure}[b]
    \center
    \includegraphics[width=0.9\columnwidth]{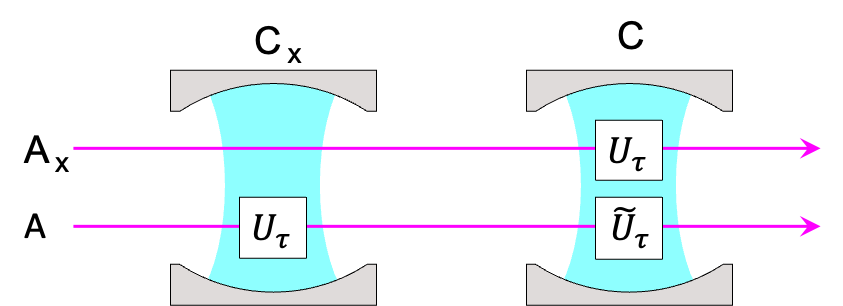}
    \caption{Schematic representation of the complete decorrelation protocol. First, the auxiliary atom \AI{} interacts with the main cavity \CII{}. Next, the main atom \AII{} interacts with the auxiliary cavity \CI{} and then, finally, with the main cavity \CII{}.}
    \label{fig:complete_decor}
\end{figure}

Since there exists no CPTP map that implements the complete decorrelation process for an arbitrary state, we experimentally simulate this process employing two atoms (\AII{} and \AI) and two cavities (\CII{} and \CI), as depicted in Figure~\ref{fig:complete_decor}. The auxiliary atom \AI{} is sent through the cavities before the main atom \AII. The atoms go through the auxiliary cavity \CI{} before entering the main cavity \CII. We can switch on and off the interactions between any atom and any cavity by detuning the atom with the corresponding Stark electric field, \setup{V} or \setup{V}$_\textrm{x}$, see~Fig.~\ref{fig:setup}. 

The implemented experimental sequence looks as follows. The auxiliary atom \AI{} interacts and gets entangled only with the main cavity \CII{}, creating the joint state $\rho_{\tau}^{A_{x}C}$ given by~(\ref{eq:rho_tau}). Next, the main atom \AII{} gets entangled with the auxiliary cavity \CI{}, creating a similar state $\rho_{\tau}^{AC_{x}}$ of the \AII{}-\CI{} system. In this part of the sequence, the atoms and the cavities are in the same reduced state $\rho_{\tau}^{A_{x}}=\rho_{\tau}^{A}$ and $\rho_{\tau}^{C_{x}}=\rho_{\tau}^{C_{}}$, respectively. Therefore, when the main atom enters the main cavity, their joint \setup{AC} state is $\tilde\rho_{0}^{\decor}=\rho_{\tau}^{A}\otimes\rho_{\tau}^{C}$, ideally reading
\begin{equation}\label{eq:rho_decor}
  \tilde\rho_{0}^{\decor} = \frac{1}{4}\left(\ket{e}\bra{e}+\ket{g}\bra{g}\right) \otimes \left(\ket{0}\bra{0}+\ket{1}\bra{1}\right).
\end{equation}
This expression illustrates the complete decorrelation applied to the entangled state $\rho_{\tau}$.

The backward process closing the cycle is implemented by the backward process $U_\text{B}$ between \AII{} and \CII{}. The system's state at the end of the cycle then reads
\begin{equation}\label{eq:rho_decor_tau}
  \tilde\rho_{\tau}^{\decor} \!=\! \frac{1}{4}\left(
  \ket{g0}\bra{g0}\!+\!
  \ket{g1}\bra{g1}\!+\!
  \ket{e0}\bra{e0}\!+\!
  \ket{\psi_2}\bra{\psi_2}\right),
\end{equation}
where the double-excitation contribution is given by $\ket{\psi_2} = \cos(\theta) \ket{e1} + \sin(\theta) \ket{g2}$ with $\theta=\sqrt{2}\pi/4$. 

The entropy production $\left\langle \Sigma_{\decor}\right\rangle$ of this process, erasing all correlations in the system, ideally reaches its maximum possible value of 2 bits. However, as in the case of the dephasing, the measured values of $\left\langle \Sigma_{\decor}\right\rangle$ and $\Delta I_{\decor}$ are significantly lower than their ideal values. The value of $\Delta I_{\decor}$ corresponds to the total mutual information in $\rho_\tau$ efficiently erased by the decorrelation, and is therefore bounded to the $0.88$ bits of information effectively prepared. The higher value of  $\left\langle \Sigma_{\decor}\right\rangle$  compared to $\Delta I_{\decor}$ can be explained in a way similar to the dephasing environment. Besides, realizing forward and backward processes with different physical systems introduces additional imperfections in the cycle. For instance, any dispersion in the physical properties, like exact position of the atoms inside the cavities, will unbalance the forward and backward parts of the cycle, realized with different atoms and cavities, thus producing extra entropy. This effect can also be partially explain the increased difference between the simulated and measured $\left\langle \Sigma_{\decor}\right\rangle$.

\subsection*{Local thermalization}

The local thermalization process in the considered model system requires resetting one of the subsystems back to its initial state. Realizing  this reset in a short time interval between $U_\textrm{F}$ and $U_\textrm{B}$ either on the atom or on the cavity is nontrivial and challenging. It demands sophisticated modification of the atom-cavity interaction sequence and inclusion of auxiliary atoms and/or microwave pulses. 

If we choose to reset the cavity state, a similar evolution can be realized by making the atom interact with two cavities, \CI{} and \CII{}, prepared in the same initial vacuum state $\ket{0}$. First, we let the atom to perform the forward evolution $U_\textrm{F}$ in \CI{}. The state of the \AII{}-\CII{} system, after tracing out the state of \CI{},  reads
\begin{equation}\label{eq:rho_reset}
  \tilde\rho_{0}^{\reset} = \frac{1}{2}\left(\ket{g}\bra{g}+\ket{e}\bra{e}\right) \otimes \ket{0}\bra{0},
\end{equation}
effectively resetting the cavity back to its initial vacuum state. The cycle is closed by the backward process $U_\textrm{B}$ between \AII{} and \CII{}, similarly to all previous protocols. Note that this experimental scheme is identical to the one of complete decorrelation, see Fig.~\ref{fig:complete_decor}, but without auxiliary atom \AI{}. The state reconstruction is finally performed on the \AII{}-\CII{} system. Ideally, the system comes to the state 
\begin{equation}\label{eq:rho_reset_tau}
  \tilde\rho_{\tau}^{\reset} = 
  \frac{1}{2}\ket{g0}\bra{g0} + 
  \frac{1}{2}(\ket{e0}+\ket{g1})(\bra{e0}+\bra{g1}).
  \end{equation}

For the sake of simplicity, the thermalization process has been realized by the same experimental sequences programmed to realize the complete decorrelation with two atoms and two cavities. Since atomic samples have random number of real atoms, the decorrelation process was implemented by experimental sequences conditioned on the final detection of both atoms, \AI{} and \AII{}. For the thermalization environment, on the contrary, we have post-selected sequences conditioned on the successful detection of only one atom \AII{} and with no atoms in the sample~\AI{}.

As in the case of complete decorrelation, the reset process ideally erases, although in a different way, 2 bits of information stored in $\rho_\tau$, \ie $\Delta I_{\reset}=2$. In contrary to other considered environments, the entropy production $\langle \Sigma_{\reset}\rangle$ is expected to diverge to infinity for the ideal reset cycle. This behavior naturally follows from the reset cavity state ($\ket{0}$), which corresponds to a classical zero-temperature state presented by rank deficient density matrix. The impossibility of preparing this state according to the third law of thermodynamics is reflected by the infinite value of the entropy production. However, any imperfection in the state reset or in the following unitary backward evolution removes this restriction making realistic $\langle \Sigma_{\reset}\rangle$ finite. But, due to the highly divergent nature of $D$, even minor fluctuations in experimental noise lead to significant random changes in the reconstructed value of $\langle \Sigma_{\reset}\rangle$, which makes this quantity less suitable for characterizing quantum systems out of thermal equilibrium.

Figure~\ref{fig:results} shows $\langle \Sigma_{\reset}\rangle$ and $\Delta I_{\reset}$. As for the dephasing environment, the erased information $\Delta I_{\reset}$ is maximal, set by the initial entangled state $\rho_\tau$. Regarding the value of $\langle \Sigma_{\reset}\rangle$, it does not correlate at all with its infinite ideal value due to various experimental imperfections and the high sensitivity of the divergence $D$ in the vicinity of rank deficient states. The discrepancy between its numerically simulated and experimentally measured values is of the similar nature as for the decorrelation protocol, also using different atoms to simulate a single qubit.

\section{Conclusions}\label{sec:conclusions}

We have simulated several types of non-thermal environment \cite{Landi2021} that erase different information and correlation in a bipartite quantum system. As a model system, we have manipulated single circular Rydberg atoms interacting resonantly with a high-finesse cavity mode. Following the two-point measurement approach, the experimental sequences are composed of three basic processes: unitary forward evolution, interaction with environment, and backward unitary evolution. The irreversibility of thermodynamical cycles containing unitary (atom-cavity interaction) and non-unitary (coupling to environment) evolution is quantified by the entropy production $\left\langle \Sigma_{\text{}}\right\rangle$, computed as the Kullback-Leibler-Umegaki divergence between the initial and final states of the system. All these states are estimated by full quantum state tomography based on quantum non-demolition measurement of the photon number in the cavity and the state-selective atom detection. 

After our initial data analysis, we encountered an unforeseen difficulty in computing entropic quantities from reconstructed density matrices. Being close to pure states (a frequent occurrence for quantum information systems) and not being full rank (the feature of the maximum-likelihood estimation), reconstructed density operators can lead to excessively strong divergent behavior of the entropy production $\langle \Sigma_{\text{}}\rangle$. The KLU divergence $D$ tends to diverge to infinity for states with different support. Being used to quantify the entropy production, this behavior of $D$ reflects the third law of thermodynamics prohibiting zero-entropy states, like \eg thermal state with zero temperature, whose preparation would require infinite entropy production. For any other classical thermal state, the density operators are always full rank, thus resulting in finite $D$. In contrast, for quantum systems in or close to pure states, like qubits in quantum superposition states, the divergence fails to adequately describe the system's evolution and its irreversibility.   

Although we have managed to partially overcome the problem of the divergence for experimental MLE states, the general question remains relevant: how adequate is the widespread use of this quantity to characterize any thermodynamical processes in quantum systems? Or should we look for a replacement that would allow us to more accurately quantify irreversibility at the quantum level with less dependence on measurement accuracy and a particular state reconstruction algorithm? We hope that this work will draw the attention of the community to this problem of quantum thermodynamics.

\section{Acknowledgments} This work was supported by the Foundational Questions Institute Fund (Grant number FQXi-IAF19-05), the Templeton World Charity Foundation, Inc (Grant No. TWCF0338), ANR Research Collaborative Project ``Qu-DICE" (ANR-PRC-CES47) and ANR Single-team Research Project ``ProRydAQS" (ANR-24-CE47-2923-01).

\appendix
\section{Monte-Carlo sampling of density matrices according to their likelihood}
\label{app_MonteCarlo}
The numerical approach to sample density matrices according to their likelihood is achieved using a standard Metropolis algorithm. First, we initialize the algorithm to an arbitrary density matrix $\rho_0$ (in practice, we start with $\rho_{\rm ML}$). We then construct a Markov chain by iterating the following steps: 
\begin{itemize}[leftmargin=1.0 cm]
    \item  propose a new density matrix $\rho_{\rm new}$;
    \item if $L(\rho_{\rm new}) > L(\rho_k)$, then accept the move;
    \item otherwise, accept the move with probability $L(\rho_{\rm new}) / L(\rho_k)$;
    \item if the move is accepted, then $\rho_{k+1}=\rho_{\rm new}$;
    \item otherwise, keep $\rho_{k+1} = \rho_k$.
\end{itemize}
Here, the index $k$ labels the iteration loops. 

A sufficient condition for convergence of the Markov chain towards the $L(\rho)$ distribution is that of detailed balance, which is guaranteed if the probability to propose the move to $\rho_{\rm new}$ starting from $\rho_k$ is the same as the probability to propose the move to $\rho_k$ starting from $\rho_{\rm new}$. This is ensured in our algorithm by proposing moves of the form $\rho_{\rm new} = \rho_k + \epsilon - {\rm Tr}(\epsilon) \frac{\mathbf{1}}{d}$, where $\epsilon$ is a $d\times d$ hermitian matrix whose coefficients are all independently drawn from a Gaussian distribution of zero mean and standard deviation $\sigma$. The dimension of the Hilbert space is $d=8$ to which we truncate our experimental analysis. Whenever $\rho_{\rm new}$ has negative eigenvalues, the move is always rejected (namely, $\rho_{\rm new}$ has a vanishing likelihood). The standard deviation $\sigma$ is automatically adjusted throughout the sampling procedure so that the mean acceptance rate of the proposed moves remains between $0.4$ and $0.6$. Finally, in order to obtain samples with a good statistical independence, between each collected density matrix we run about $10^5$ individual Metropolis steps as described above, and collect in this way an ensemble of $n=100$ density matrices for each decorrelation scenario. 

\bibliography{biblio.bib}

\end{document}